# A Structural Parameter for High Tc Superconductivity from an Octahedral Möbius Strip in RBaCuO:123 type Perovskites

R. Pérez-Enríquez
*Deptartamento de Física,
Universidad de Sonora*
Marzo de 2001.

**Resumen.** Se introduce un índice para la caracterización de los superconductores de tipo Perovskita. Este índice, denominado Parámetro Estructural de la Superconductividad de Alta Tc (PESATc), se calcula mediante el uso de un Orbital de Möbius definido como la longitud de la Banda de Möbius Octaédrica sobrepuesta a la estructura en forma de octaedro formada por los planos de CuO y los Oxígenos apicales, O(4). Se presentan los resultados obtenidos para compuestos YBCO:123 cuyas posiciones atómicas que están reportadas en la literatura, fueron medidas por medio de Difracción de Neutrones y de Rayos-X. Se reporta el hallazgo de una dependencia de la Temperatura crítica con el valor del parámetro estructural; además, se encuentra que, para un conjunto de compuestos del tipo RBaCuO, la dependencia entre el PESATc y la Temperatura crítica es lineal. En el primer caso, se encuentra un intervalo de valores para el PESATc en el cual se presenta el comportamiento superconductor; asimismo se obtiene que el valor correspondiente a la máxima Tc está bien definido. En el segundo caso, el valor de la pendiente, obtenida para un conjunto de siete compuestos de Tierra Rara, es de 4.515 con un coeficiente de correlación de 0.9402. El único caso que se alejó de la tendencia general fue el compuesto con base al La; la razón pudiera ser el alto contenido de Oxígeno, mayor a 7.

**Abstract.** An index for the characterisation of the Perovskite type Superconductors is introduced. The index, denominated as Structural Parameter of High Tc Superconductivity (PESATc from its name in Spanish), is calculated using a Möbius Orbital defined as the length of an Octahedral Möbius Strip overlaying the octahedra created by the CuO planes and the apical Oxygen atoms, O(4). Results for structures of YBCO:123 type compounds are presented; their atomic positions were measured by neutron diffraction and x-ray diffraction techniques as reported in the literature. It has been found that critical temperature depends on the value of the measured structural parameter; furthermore, for a set of RBCO compounds there is a lineal response between Tc and PESATc. In the former case, a specific range for the superconducting behaviour is established by PESATc values, and a well-defined point for the maximum Tc is found. In the latter one, a slope of 4.515 with correlation coefficient of 0.9402 was found when seven Rare Earth compounds were considered. The only compound that goes out from this trend, is the La based compound; the reason could be the greater than 7 Oxygen content.

## Introduction

Since the discovery of Perovskite type compounds [1], High Tc Superconductivity had and extraordinary development. It seemed that the way to reach critical temperatures of the order of that of dry ice (solid $CO_2$) had been conquered. However, all results on critical temperature that grew in just five years, have become staged mainly because a theory for the behaviour of the ceramics has not yet come. In 1990, V. Tokura and T. Arima [2] did introduce the Block Layer concept (BL) which combined with the CuO planes, allowed them to make a fairly complete classification of Superconducting compounds. With the aid of this classification, they tried to design new compounds.

Between most successful Perovskite structures, those containing a great number of CuO layers ($CuO_2$) were those which render best results while searching for the highest critical Temperature. So, the way of searchers took one oriented to build multilayered structures of this kind; they search for them no matter the means required to make them stable; it was used high pressure or mechanical confinement of several compounds with a fragile stability. To some extent, the perovskite's way was fruitful but without reaching the objective they looked for.

Furthermore, in superconductors with CuO planes, the critical temperature had risen beyond 130 K. Families of YBCO, BSCCO, TBCCO and HgBCCO compounds have their maximum Tc at 95, 110, 125 and 133 K, respectively [3]. Other perovskite structured ceramic without the mentioned cuprates have also superconductor behaviour like the SrRuO with Tc at .93 K and, a just discovered non-oxygen one, the MgCNi which critical temperature is 8 K [4, 5].

In this process, the attention of the theoretical physicists and experimental ones has been centred on the analysis of the Copper-Oxygen layers (CuO) and the role of the Oxygen content on the structure. As a consequence, it has been attempted to find some kind of structural parameter, which would allow represent the conditions needed to reach a superconducting state in these types of materials. One such a parameter is named ASIN (Asymmetry Index)[6] and is widely used. Furthermore, it has been proposed that high critical temperature superconductivity has its origin, mainly, in a bidimensional charge transfer process (through the CuO layers).

However, since middle 80's, an increasing interest in the role of an interlayer coupling has been developed. In 1987, Zlatko Tesanovic [7] made an analysis about the origin of electron pairing in these perovskite type ceramics. In his review, Tesanovic put emphasis on a possible Josephson effect mechanism between layers as the mean to provide stability and enhancement of Tc. On the other hand, while studying the role of oxygen content in the orthorhombic structure, it has been found that the interactions between layers and chains must be involved in the original mechanisms of superconductivity [8]. Furthermore, in a very good analysis on how deformation of YBCO ceramics has a definitive influence in their mechanical properties was made by VS Bobrov in 1993[9]. Again, he puts special interest on the influence of oxygen content on its above mentioned characteristics.

These assumptions made along the years that followed the discovery of the High Tc Superconductors, could be considered included while an analysis of the *dogmas* introduced, recently, by Anderson in his "Theory of Superconductivity in The High Tc Cuprates" [10]. In summary, a careful reading of them, putting special attention to those related with the role of CuO2 planes and the two-dimensional transport of charge in combination with the interlayer hopping can support the proposition of a structural parameter involving these facts.

With these ideas in mind, the Structural Parameter of High Tc Superconductivity (PESATc from

'Parámetro Estructural de la Superconductividad de Alta Temperatura crítica') is proposed and used for the study of specific compounds [11]. The YBCO was first analysed and later on, compounds on which Yttrium atoms have been replaced by Rare Earth ones. This could be done because the observed behaviour of layers and apical Oxygen atoms, O(4), was the same, rendering to the following: a charge pumping from CuO chains[12] to CuO planes allow to think on the possibility that Cooper pairs (mediators of superconductivity) evolve in space through those regions. Then, PESATc index is based in the evaluation of a relation between the length of a Octahedral Möbius Strip overlying the Perovskites octahedron and the length of the Cu(2)-O(4) link.

In order to present the PESATc index, a brief summary of Perovskite type ceramics is done, putting special attention on YBCO:123. With this background, a simple description of the octahedral representation of the Möbius Strip is made, which allows presenting an overlying Octahedral Möbius Strip on perovskite unitary cell. At the end, making use of this model and with some structural data reported in the literature, PESATc index is calculated for 1:2:3 Perovskite type of compounds. In order to show the usefulness of this PESATc index, results for the following cases are presented:

a) case of YBCO:123, the dependence in PESATc values in accordance with the changes in Oxygen content and the substitution of Y by Pr are evaluated [16]; and,
b) case of RBCO compounds, the Tc as a function of the R element (R= Dy, Y, Gd, Sm, Ho, Eu, Er, La) is analysed. [17-23].

Finally, some comments on how the model could be applied to other superconducting compounds, which accept the introduction of some kind of Octahedral Möbius Orbital or other kind of itinerant electron paths, are given.

## The YBCO:123 and Its Perovskite Type Structure

The Superconductor ceramic of YBCO:123 was the first Perovskite structured ceramic which presented a superconducting state with critical temperature above that of liquid Nitrogen ($N_2$). With a critical Temperature of 92 K, it defined a search for High Tc Superconductivity fever. Because of it, this is the most studied compound until today.

The ideal structure of Perovskites has a general equation $ABX_3$; being A and B metallic cations, while the third element, X, is a non-metallic ion. In the general case, crystals of this type of compound grow in cubic or octahedral arrays like the Calcium Titanate ($CaTiO_3$)[1]. It is important to say that the electric properties between these materials change in agree with the size of the ions involved in the structure; they induce deviations on the shape of the octahedral arrays. Such is the case of the artificial Perovskite $BaTiO_3$ (Barium Titanate) which ferroelectric behaviour is very important; being a typical example of decentered cations.

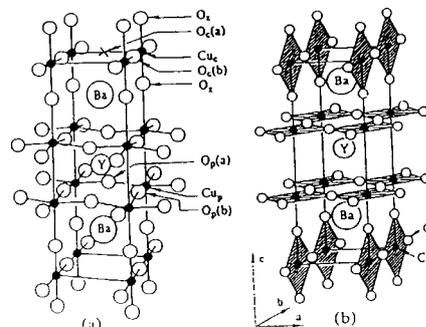

Figure 1. Structure of Orthorhombic $YBa_2Cu_3O_{7-y}$. a) Atomic positions, b) Copper planes and chains

In the case of $YBa_2Cu_3O_{7-y}$, the ideal structure of perovskite type suffers several transformations because of the presence of Y ions located at the Ba positions. Then, in this kind of compounds, the unitary cell becomes from three normal cells stacked as it is shown in Figure 1a. As can be seen on Figure 1b, depending upon the occupancy of the Oxygen planes, the structure presents CuO layers and CuO chains. The planes continue to form the octahedral structures when apical Oxygen atoms, O(4), are taken into account. If little modifications to the structure causes severe changes in the electric, optical and elastic behaviour in those materials [1], it became evident that a way to measure the deviations form this structure is needed; i.e., a way to evaluate the perovskite structure change in this compounds must be defined.

## The Octahedral Representation of Möbius Strip

The Möbius Strip is a surface with only one side and only one face. There is an easy way to build it from a paper strip or band: it is enough to take the strip and glue its ends, having the precaution to roll one of them by 180 degrees before joining. However, its octahedral representation is difficult to visualise.

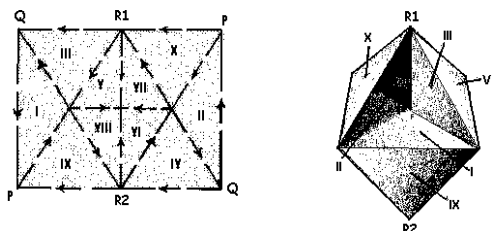

Figure 2. Octahedral representation of Möbius strip: a) Triangulation of Möbius Strip, b) Octahedral Representation.

Möbius Strip has several properties. The most relevant to the analysis here presented, is that one of its non-orientability; this characteristic makes it seem like the behaviour of electron spin. If a vector **n** perpendicular to the plane of the strip is taken into account at any point **p**, it will change its orientation while travelling along the strip by its central line. As it comes back to the **p** point a –**n** vector appears. In order to obtain the original orientation of the vector, it is necessary to make another turn over the strip. This is why is said that the Möbius Strip is a non-orientable bidimensional manifold [13].

This topologic space, similarly to other bodies and surfaces, accepts several representations. When a Möbius Strip is divided by polygons, it could be marked by triangles with and Euler characteristic equal to $N_0 - N_1 + N_2 = 0$, with N values of 7, 17 and 10, respectively. In Figure 2a, a rectangular mapping could be appreciate and, in Figure 2b, its development by triangles which allows to build a 3-dimensional octahedral representation.

As it is observed in Figure 3, the octahedral form is built by four isosceles triangles I, II, III and IV (with a unitary Hypotenuse) and by six equilateral triangles V, VI, VII, VIII, IX, and X (with a unitary side length). Following the procedure of bending the strip's end sides, the joint between triangles I and II obeys the half turn condition before gluing; obtaining in this way the non-orientable surface.

As it was shown earlier, one of the specific properties of the Perovskite structure is that related with the identification of Oxygen atoms around the cationic ions as a shape with octahedral symmetry (see Figure 1b). This fact allows making an association of such structure with a Möbius Octahedral Strip as it has been developed until now.

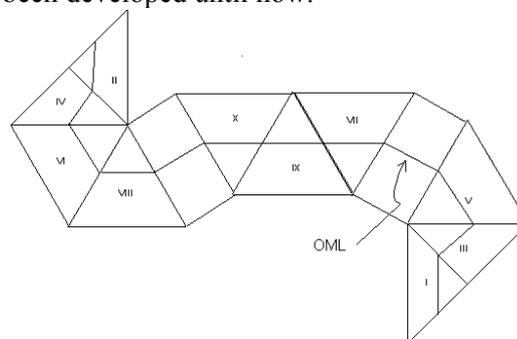

Figure 3. Octahedral Möbius Strip model modified to overly in Perovkite ceramic structure. The centre line shows the path to calculate OML.

## Octahedral Möbius Length

The Octahedral Strip could be modified with two small rectangles in order to get the shape shown in Figure 3. This new modified strip could be overlaid in the Perovskite's octahedral structures of the typical superconducting compounds (see Figure 2b). The rectangles needed are: one to be join between V and VII triangles;

another between VIII and X. Such a strip and the way it could be done, makes it possible to incorporate surfaces of this kind to the Octahedral Perovskites which are typical of the superconductor compounds.

Once the Octahedral Möbius Strip is overlying the perovskite structure, a measure of its length will render a factor that represents the specific compound. As it will be show later, the behaviour of this Octahedral Möbius Length (OML) depends on the characteristics of the ceramic being considered as well as its amount of Oxygen content. However, there is any special relation observed between OML and Tc.

Möbius Strip and the Perovskite the sides of rectangles are given by cell parameters as follows: $a$ by CuO to CuO planes. And sides of the squared rectangles I and II: CuO to O(4) length.

In Table I, OML calculations for two sets of samples are shown. Both sets comes from samples of YBCO:123 ceramics: in the first one (A1..H1), the lengths considered are from samples prepared by different ways [12], reflecting a variable oxygen content (y = 0, 0.2, 0.4, 0.6, and 1); the second set, (A2..E2), belongs to the $Y_{1-x}Pr_xBa_2Cu_3O_{7-d}$ compound in which Ytrium was gradually replaced by

| TABLE I. Octahedral Möbius Length for YBCO Type samples | | | | | | | | |
|---|---|---|---|---|---|---|---|---|
| Sample | A1 | B1 | C1 | D1 | E1 | F1 | G1 | H1 |
| OML | 20.698 | 20.731 | 20.753 | 20.774 | 20.748 | 20.787 | 20.823 | 20.848 |
| Tc | 71 | 61 | 48 | 30 | 15 | 0 | 0 | 0 |
| Sample | A2 | B2 | C2 | D2 | E2 | | | |
| OML | 20.653 | 20.698 | 20.752 | 20.796 | 20.991 | | | |
| Tc | 92 | 76 | 48 | 0 | 0 | | | |
| Notes. Samples 1 for Yba2Cu3O7-y with different Oxygen content.[4]      Samples 2 for Y1-xPrxBa2Cu3O7-y with x= 0, .2, .4, .6, .8 [13] | | | | | | | | |

In this section, the values for the OML for 1:2:3 Perovskite type compounds is presented: firstly, YBCO ceramics with changing structure due to both Oxygen content and Yttrium replacement by Praseodymium is analysed; and, secondly, substitution of Yttrium by other Rare Earth elements is considered. In all cases, data structural parameters considered were coherent, the former coming from neutron diffraction technique and the later based on X-ray diffraction results.

## Octahedral Möbius Length for YBCO Ceramics

The Octahedral Möbius Length could be calculated with precision, if enough information about the position of each atom on the structure is known, in such a way that the interatomic distances can be estimated. The OML is the length of the path that goes through the middle of the octahedral representation shown in Figure 3. When making the association between

Praseodymium, values for x = 0, 0.2, 0.4, 0.6 and 0.8 [16]. It is important to make emphasis in the fact that structure's modification induced by the Pr atoms causes lose of superconducting behaviour in a similar way the defect of Oxygen does.

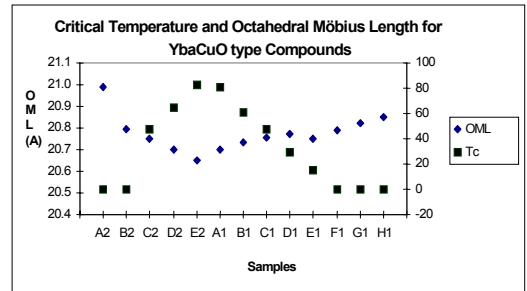

Figure 4. OML and Tc for YBCO:123 Compounds.

The Figure 4 presents a combination of the critical Temperature and Octahedral Möbius Length (Tc and OML, respectively) for those samples. As it can be seen, when considered as a whole, the Tc for them has well-defined trends; coming from the non-superconducting states for samples with defect in Oxygen atoms and a high Pr content, to increasing and decreasing trends towards and from a maximum Tc for the typical $Y_1Ba_2Cu_3O_7$ ceramic. Although there is a special behaviour of the OML, corresponding trends are not defined as in the Tc case. However, an anticorrelation between them can be established.

# The Structural Parameter of High Tc Superconductivity

Before the main proposal of this paper, it is useful to compile in few words the lines of thought followed until now and which will support the introduction of the Structural Parameter of High Tc Superconductivity

| TABLE II. Octahedral Möbius Length* and PESATc* for RBCO Compounds (R = Rare Earth) | | | | | | | | |
|---|---|---|---|---|---|---|---|---|
| Rare Earth | Dy | Y | Er | Ho | Eu | Sm | Gd | La |
| OML | 20.829 | 20.719 | 20.721 | 20.560 | 20.953 | 20.982 | 20.825 | 21.373 |
| PESATc | 8.91 | 8.923 | 9.013 | 9.219 | 9.274 | 9.418 | 9.418 | 9.740 |
| Tc** | 92.0 | 91.3 | 92.5 | 92.9 | 93.9 | 93.8 | 94.0 | 80.0 |
| *Measured from X-ray diffraction data, [14-20]. | | | | | | | | |
| **Middle point of critical Temperature. | | | | | | | | |

## Octahedral Möbius Length for RBCO:123 ceramics

Following the same methodology for the calculation of OML, the Table II was prepared. On it, the results for samples analysed by the same research group were used. In fact, all structural parameters used come from articles published by the Materials Science Institute Group, of the University of Tsukuba, Japan, in JJAP in 1987 [18-23]. In this case, calculations were made for the superconducting item for RBaCuO samples (where R=Dy, Gd, Er, Ho, Y, Sm, Eu, La). The Figure 5 shows how the OML behaviour is not as defined as in the later case analysed. One remarkable thing to be considered is the way the OML for La compound come out of them.

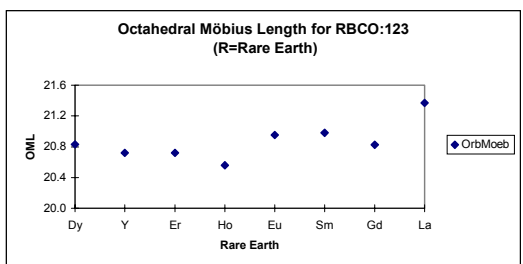

Figure 5. OML for samples with Y substituted by a Rare Earth.

(**PESATc** from the expression **P**arámetro **E**structural de la **S**uperconductividad de **A**lta **T**emperatura **c**rítica).

The structure of ceramic compounds as perovskite type superconducting ceramics is essential for its mechanical and electrical behaviour. In most of them, the octahedral of the perovskite is fundamental; such as the way the embedded octahedra are distorted or modified but not broken. Then, an overlying Octahedral Möbius Strip could be inserted and its length could be measured (OML). Now, a specific structural element must be added in order to characterise in a better way the structure (see Figure 6.).

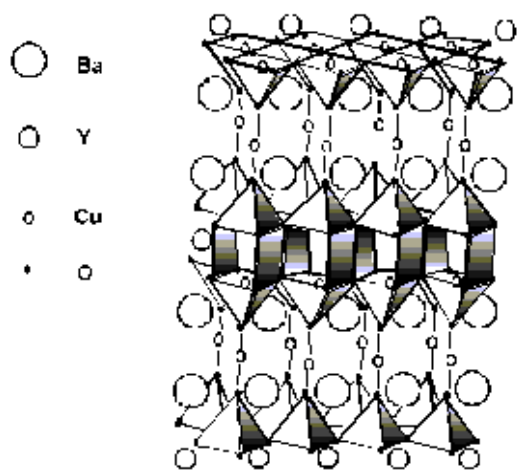

Figure 6. Diagram showing how the Octahedral Möbius Strips overlay in YBCO structure.

For this purpose, it is necessary to remember the considerations made by the

*dogmas* introduced by Anderson [7]. One of them, considered already in the OML, is the role of apical O(4) atoms, in both the chain formation and the charge transport to the planes. With this idea in mind, a convenient characteristic to be included in such a structural parameter for superconductivity could be that of the bonding distance between Cu(2) and O(4) (Cu-O(4) Bond); mainly, because this could be seen as a measure of how much squeezed the octahedrons are. Then, the PESATc is defined taking into account both aspects as shown below. This new structural parameter for these types of superconductor materials could be useful as will be clear from the following paragraphs:

$$PESATc = \frac{Octahedral\ M\ddot{o}bius\ Length}{Cu - O(4)\ Bond}$$

the evaluation of Oxygen content could help in designing process for new ceramics.

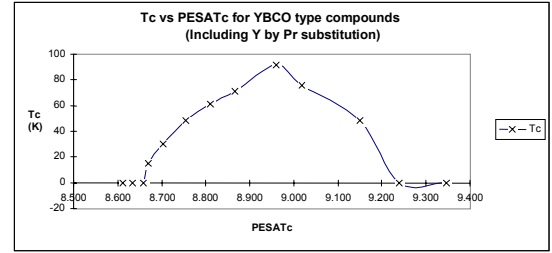

Figure 7. Dependence relation between Tc and PESATc.

## PESATc for RBCO:123 Superconducting Perovskites

When PESATc values are calculated for crystals compounds on which Y has been replaced by a Rare Earth atom (the above eight elements), the trend of Tc vs PESATc becomes a straight line as can be seen in

| TABLE III. PESATc* and Tc for YBCO type compounds | | | | | | | | | | | | |
|---|---|---|---|---|---|---|---|---|---|---|---|---|
| Sample | A2 | B2 | C2 | D2 | E1 | A1 | B1 | C1 | D1 | E1 | F1 | G1 | H1 |
| PESATc | 9.35 | 9.24 | 9.15 | 9.02 | 8.96 | 8.87 | 8.81 | 8.75 | 8.70 | 8.67 | 8.66 | 8.63 | 8.61 |
| Tc | 0 | 0 | 48 | 76 | 92 | 71 | 61 | 48 | 30 | 15 | 0 | 0 | 0 |
| * Calculated from Neutron Diffraction data [4, 13] | | | | | | | | | | | | |

## PESATc for YBCO Ceramics

In Table III, a summary of PESATc index values for YBCO ceramics is presented. As can be seen in the Table, the index has a monotonic decreasing trend with values going from 9.346 to 8.611 for the above mentioned set of samples analysed by neutron diffraction. This fact allows make the graph shown in Figure 7. From it, general trends of the dependence of Tc with PESATc could be obtained: a) it defines a range of values for which the compound is not a superconductor; b) it has a corresponding value of 8.96 for maximum Tc in the YBCO compound; and, c) in the case for Oxygen content samples values (A1..H1), the curve obtained shows a smooth behaviour different from that obtained when Tc is directly graphed vs the Oxygen content [21]. This enhancement in

Figure 8, where values from Table II are graphed. The linear estimate reaches a 10.130 value when La compound is taken out of consideration. The slope of the straight line is m = 4.155, obtained with a correlation coefficient of R = 0.9402 which gives a very good confidence. The reason why Lanthanum (La) based compound goes out of the general trend is not clear but could be related with the above 7 of Oxygen content (7.62 for the Orthorhombic phase) [20].

## Other structured materials

The importance of perovskite structure or, more precisely, the presence of atoms arranged in a way that an octahedron could be visualised, seems to be a constant in new superconducting materials. Several compounds found in last few years show this fact ($MgCNi_3$, $MgB_2$ and $Sr_2RuO_4$).

Those structures do not present critical temperatures as high as those found on CuO perovskites but all of them have associated itinerant electrons that could be involved in a kind of the proposed Möbius Orbitals that give rise to the PESATc index. Overly octahedron needs to be allocated between the atoms in the structure and its length measured. In other materials like C60 or manganetites, the search for some kind of Möbius path is the way to follow.

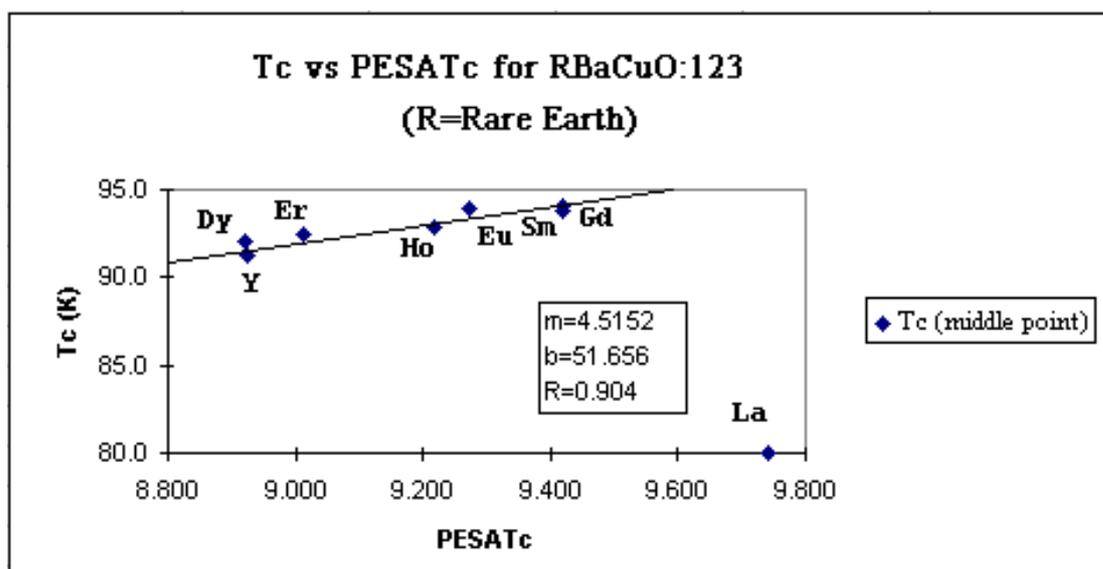

Figure 8. Tc and PESATc show a linear dependence when different Rare Earth atoms are used (see text).

## Conclusions

The presence of Perovskite type structures on High Tc Superconductors allows the overlay of Octahedral Möbius Strips on the unitary cells of these ceramics. The path along such a overlying strip could be measured with the aid of diverse diffraction techniques information (neutron diffraction and x-ray diffraction); obtaining by this mean a Octahedral Möbius Length (OML) representing that structure. Motivated by the role played by O(4) in the chain formation and charge pumping to the copper planes, a index is calculated by the ratio between OML and the CuO-O(4) Bond. This index has been called PESATc after its name in Spanish: Parámetro de la Superconductividad de Alta Tempratura crítica. It shows to be useful for the analysis of the change in structure related to the critical Temperature. In this case the curve of Tc vs PESATc who's Oxygen content changes shows an increasing trend better defined than that obtained by Tc vs Oxygen content. For YBCO compounds, there is a range for superconductor behaviour going from 8.66 to 9.24; showing a maximum Tc = 92 K around 8.9 of PESATc.

A linear dependence of Tc with PESATc for RBCO:123 compounds have been found. These results are consistent for seven Rare Earth elements (Dy, Gd, Er, Ho, Y, Sm and Eu) and show a slope of 4.155 with a correlation coefficient of 0.94. The only case beyond this trend is the Lanthanum based ceramics.

Both results give a fairly confidence on the PESATc index as a structural parameter useful for the study of Perovskite type Superconductors. Superconducting behaviour in other perovskite structured compounds, including those without cupprates, and on other complex structured materials like buckminsterfullerenes, magnesium diboride and manganetites, suggest that further work has to be done in order to show PESATc's generality.

## Acknowledgements

I want to express my must grateful thanks to all people that in some way supported me to get this paper done. I have to recognise specially Dr Roberto Escudero, IIM-UNAM, Dr Raul Riera, DIFUS, and Dr Malcolm Raven, DEEE-U.Nottingham, for orientation and discussions. Also, I have to mention that this would not be finished without the PROMEP and The University of Nottingham supports.